\documentclass{an}
\usepackage{graphicx}
\usepackage{times}
\usepackage{fancyhdr}
\begin{document}

\title{Diffuse Radio Sources in Groups and Poor Clusters}

\author{Kisha M. Delain \and Lawrence Rudnick}
\institute{University of Minnesota, 116 Church St SE, Minneapolis, MN  55455}

\date{Received $<$date$>$; 
accepted $<$date$>$;
published online $<$date$>$}

\abstract{We have discovered new diffuse radio sources likely associated with groups of galaxies at low redshift (0.01-0.04) and without apparent AGN by using the WENSS and WISH catalogs to perform an unbiased survey.  These sources resemble the radio halos, mini-halos, and 'relics' of rich clusters, which are thought to be powered by shocks and turbulence from infall into their deep potential wells.   Our detection of similar sources within the shallow potential wells of groups of galaxies challenges this model. Their radio luminosities are approximately two orders of magnitude higher than expected from the extrapolation of the apparent rich cluster radio/X-ray luminosity relation.  Even if these sources are misidentified distant clusters, they would lie above the apparent rich cluster radio/X-ray luminosity relation in the literature, suggesting that detections of radio halos and relics thus far may be more biased than previously thought.
}

\correspondence{kdelain@astro.umn.edu}

\maketitle

\section{Introduction}

Various selection effects complicate the interpretation of the observed radio vs. X-ray luminosity correlations for rich clusters.  At high X-ray luminosities, the upper limits fall just below the observed correlation, so deeper observations could weaken the correlation considerably.  Conversely, since radio searches have concentrated on rich clusters, a population of low $L_X$ but high $L_R$ could be missed.  Our unbiased search, described in a companion paper (Rudnick, Delain and Lemmerman 2005), of the entire sky covered by the Westerbork Northern Sky Survey (WENSS) at 327MHz begins to search for diffuse radio emission in low-mass systems.

The WENSS covers the sky north of $\delta=30^\circ$ with an angular resolution of 54'' by 54'' csc $\delta$.  The publicly available images have a typical noise value of 3.6 mJy/beam.  Instead of a targeted search, we used a multiresolution filtering technique (Rudnick 2002) to decompose WENSS into structures on different scales.  The resulting ``open'' maps contain the large scale structure greater than 3' up to 1$^\circ$, while the filtered maps contain the small scale structure.  Open maps were then searched by eye for diffuse structures, resulting in a wide variety of known sources (e.g. SNRs, HII regions, radio galaxies, known cluster halos/relics).  At least five of the brightest sources had no clear identification.  Here we discuss the three diffuse radio sources which may be associated with groups or poor clusters of galaxies.

\section{Results}

The three groups/poor clusters of galaxies identified so far do not follow the cluster $L_R-L_X$ relation (see Figures 1 \& 2).

\begin{figure}[h]
\resizebox{\hsize}{!}
{\includegraphics[]{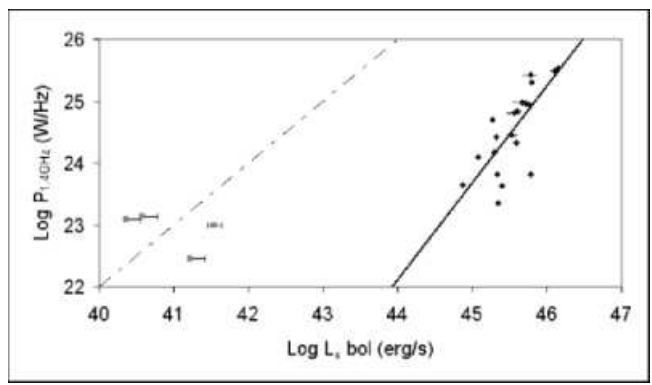}}
\caption{Radio power at 1.4GHz vs. X-ray bolometric luminosity.  Rich cluster "relics'' are shown with circles along with the best fit line to the cluster "halo'' data (Bacchi et al. 2003) extrapolated down to lower luminosities.  Squares are the groups/poor clusters, with upper limits on the X-ray data except for a detection of 0914+30. The two distinct pieces of 0809+39 are plotted separately.  The dashed line shows the dependence on the assumed redshift for the diffuse sources -- even if these sources are misidentified cluster relics, they would not fall on the previously-observed correlation.}
\label{label1}
\end{figure}

\begin{figure*}
\resizebox{\hsize}{!}
{\includegraphics[]{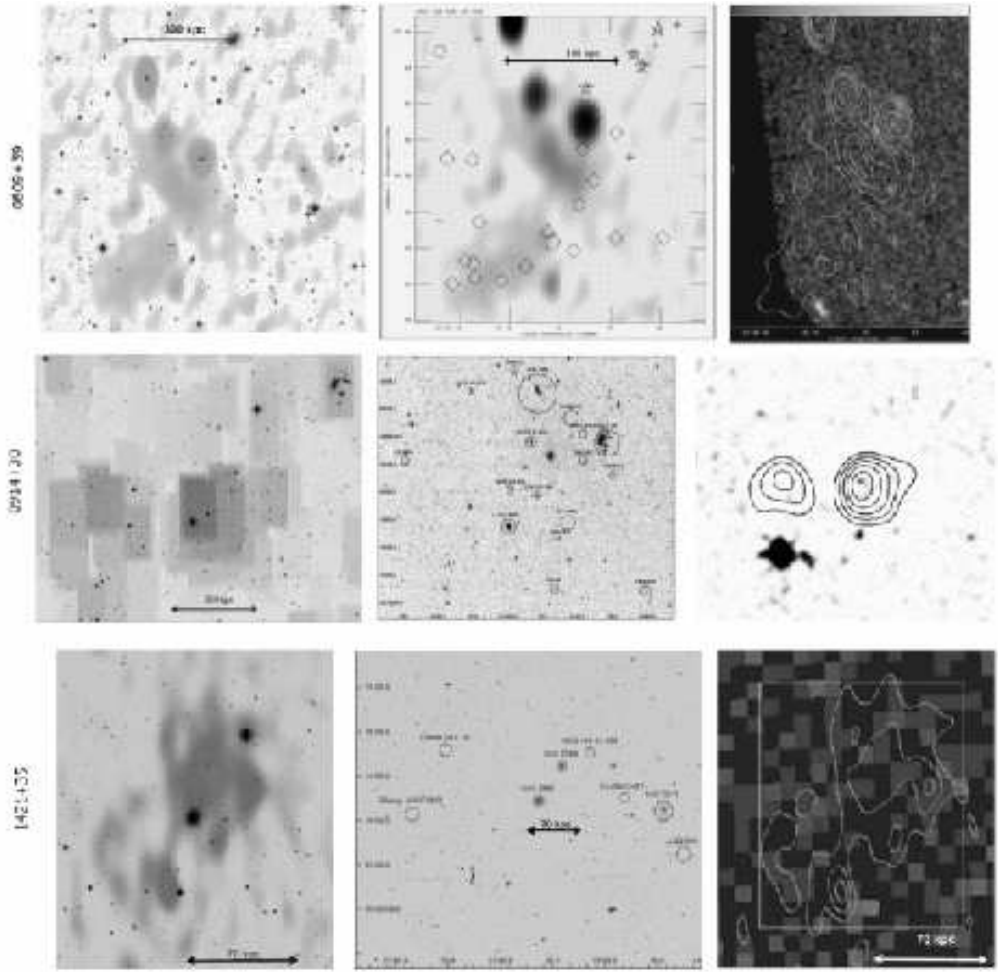}}
\caption{{\it Left:} Unfiltered WENSS with DSS galaxies for 0809+39 and 
1421+35; filtered WENSS with DSS for 0914+30. {\it Middle:} DSS image with 
groups and galaxies at group redshift. {\it Right:} XMM greyscale with WENSS 
contours for 0809+39; ROSAT greyscale with WENSS contours for 0914+30 and 
1421+35.  Comments on individual sources:  
{\bf 0809+39:} z=0.04, P$_{1.4}\sim10^{23.1}$ W/Hz, L$_X\sim10^{40.4}$ erg/s 
and $L_X<10^{40.6}$ erg/s for the two pieces (plotted separately in Fig 1).  
There are three bright background AGN in addition to the poor cluster. Diamonds show galaxies at z=0.04, stars show galaxies at z=0.075.
No clear X-ray detection is seen.  {\bf 0914+30:} z=0.02, 
P$_{1.4}\sim10^{23}$ W/Hz,  L$_X\sim10^{41.5}$ erg/s.  Weak X-ray emission is seen coincident with center of right radio contours.  Bright X-ray source to the south of the left radio emission is a background quasar.
HCG 37 shows evidence of merging galaxies and AGN activity but is 400kpc away 
from the diffuse radio emission.  {\bf 1421+35:} z=0.01, P$_{1.4}\sim10^{22.4}$ 
W/Hz, L$_X<10^{41.2}$ erg/s.  No X-ray emission is detected.}
\label{label3}
\end{figure*}

\section{Conclusions}

We find that poor clusters and groups can have diffuse radio 
luminosities comparable to those of some rich clusters, placing them two 
orders of magnitude above the cluster L$_R$-L$_X$ relation.  The sources 
are similar in size ($\sim$~few hundred kpc) to mini-halos but lack 
their corresponding dominant central galaxies and luminous X-ray cooling 
flows.

As low-mass systems are important for cosmological reasons, it is vital 
to continue searching for radio emission in groups.  Whether or not 
these are relic radio galaxies, or a new class of diffuse radio objects 
or involve similar processes as in rich clusters, our ideas of particle 
acceleration and magnetic fields will change.

\acknowledgements
The Westerbork Synthesis Radio Telescope is operated by the ASTRON (Netherlands Foundation for Research in Astronomy) with support from the Netherlands Foundation for Scientific Research (NWO).  The Digitized Sky Surveys were produced at the Space Telescope Science Institute under U.S. Government grant NAG W-2166.
Archival observations obtained from XMM-Newton, an ESA science mission with instruments and contributions directly funded by ESA Member States and NASA, and ROSAT archives from HEASARC.  Partial support from the US NSF grant 03-07600 to the University of Minnesota.\\

\end{document}